\documentclass[aps,pra,twocolumn,showpacs,floatfix,groupedaddress]{revtex4-1}
\usepackage{eurosym}
\usepackage{graphicx}
\usepackage{subfigure}
\usepackage{color}
\usepackage{soul}
\usepackage{textcomp}
\usepackage{bigints}
\usepackage{amsmath}

\begin{document}

\title{Self-Regulated Transport in Photonic Crystals with Phase-Changing Defects}

\author{Roney Thomas$^{1,*}$, Fred M. Ellis$^1$, Ilya Vitebsky$^2$, Tsampikos Kottos$^1$}

\address{$^1$Department of Physics, Wesleyan University, 265 Church St. Middletown, CT, 06457}
\address{$^2$Air Force Research Laboratory, Sensors Directorate, Wright-Patterson Air Force Base, OH-45433}
\address{$^*$Corresponding author: rthomas03@wesleyan.edu}

%
%
%







\begin{abstract}
Phase changing materials (PCM) are widely used for optical data recording,
sensing, all-optical switching, and optical limiting. Our focus here is on
the case when the change in the transmission characteristics of the optical
material is caused by the input light itself. Specifically, the
light-induced heating triggers the phase transition in the PCM. In this
paper, using a numerical example, we demonstrate that incorporating the PCM
in a photonic structure can lead to a dramatic modification of the effects
of light-induced phase transition, as compared to a stand-alone sample of
the same PCM. Our focus is on short pulses. We discuss some possible applications of such phase-changing
photonic structures for optical sensing and limiting.
\end{abstract}

\maketitle

\section{Introduction}

Phase-changing materials (PCMs) have been used in a variety of different applications in optics ~\cite{Forrest, Lieber, Wang, Basov07, Zheludev, Yang}. 
A typical optical PCM can be reversibly switched between two phases with different refractive indices, optical absorption, or electrical conductance. The 
phase change can be caused by the input light itself (self-induced phase transitions) due to heating or some other physical mechanisms~\cite{Cavalleri,
Morin,Basov11}. Alternatively, the phase transition can be induced by an external heating or cooling, by application of electric or magnetic field, or by 
mechanical stress~\cite{Dressel, Maa, Locquet98}. In this paper, we exclusively focus on the effects of self-induced phase transition in a PCM.

There are at least two different kinds of self-induced phase changes. An example of the first kind of PCMs is presented by germanium-antimony-tellurium 
alloys undergoing amorphous to crystalline phase transition when subjected to laser irradiation~\cite{Friedrich00}. The two solid phases of this alloy have 
different refractive indices, but they are both stable at room temperature. The phase transition can be reversed by a laser pulse of different duration, or 
by application of a dc electric field. This and similar systems revolutionized the media and data storage industries. A qualitatively different kind of optical 
PCMs is presented by vanadium dioxide ($VO_{2}$)~\cite{Kats13, Basov15, Goodenough71}. This material undergoes a metal-dielectric phase transition 
just above room temperature~\cite{Averitt12, Leitenstorfer}. In this case, only one of the two phases is stable at any given temperature (with the 
exception of a small hysteresis area). The dielectric phase is only stable below the transition temperature $T_{C}$, while the metallic phase is stable above $T_{C}$. The electric conductivity of the metallic phase is higher than that of the dielectric phase by 3 to 5 orders of magnitude~\cite{Morin,Averitt12,Kats12,
Hashemi, Junqiao,Kwok}. The light-induced change from dielectric to metallic state is due to heating. When the temperature drops below $T_{C}$, the 
dielectric phase is restored and the material becomes optically transparent. Optical PCMs of the second kind can be used in all-optical modulators and 
switches, IR sensors, and optical limiters. The focus of our study is on PCMs similar to VO$_{2}$, in which only one of the two phases can be stable at 
any given temperature. Since the two solid phases have different material properties, their optical characteristics will also be very different before and 
after the phase transition. This difference is the basis for all known practical applications of PCMs in optics.

In this study, we go further and investigate what happens if the PCM is not just a stand-alone layer (SA) (such as a film on a substrate), but a part of a multilayer
photonic structure. Specifically, a PCM can be a defect or a chain of defects inside a photonic band-gap structure. Incorporation of the known PCMs 
into judiciously designed photonic structures can dramatically change optical manifestations of the light-induced phase transition. Specifically: 1) it 
can qualitatively modify the change in the transmission characteristics of the photonic structure associated with the phase transition in its phase-
changing component; 2) it can significantly change-- either increase, or decrease -- the critical value of the input light intensity triggering the phase 
transition; and 3) it can protect the PCM from the heat-related damage by shielding it from the high-intensity input light after transition to the high-temperature 
phase has occurred. The above effects are very much dependent on the pulse duration and the thickness of the PCM layer. This paper presents the results 
of time-domain numerical simulations for the case of short input pulses with the fluence large enough to trigger the dielectric-to-metal phase transition 
in the PCM. We compare the results for a stand-alone PCM layer with those for the same PCM layer incorporated in the multilayer photonic structure. We show 
that the judiciously designed photonic structure can: 1) drastically reduce the critical value of the input light fluence (hereinafter, the input fluence 
threshold) triggering the onset of the dielectric-to-metal phase transition in the PCM; 2) reduce by orders of magnitude the transmittance above the 
input fluence threshold, while rendering the layered structure highly reflective; 3) significantly reduce the field intensity in the vicinity of the PCM layer 
above the input fluence threshold, thereby, preventing the PCM layer from overheating; and 4) significantly enhance the field intensity in the vicinity of the 
defect layer below the input fluence threshold. The last feature relates to low-fluence pulses; it has nothing to do with the phase transition. Still, it can 
be useful for resonant sensitivity enhancement of the sensors (e.g., micro-bolometers \cite{bolometer}) located in the vicinity of the defect layer. By 
contrast, the first three features relate to high-fluence pulses; all three of them can be particularly attractive for optical limiting and switching.

As a numerical example, we consider a periodic layered structure with vanadium dioxide as a phase-changing defect layer. This simple arrangement
allows us to illustrate all the features listed above. We choose the mid-infrared (MIR) domain (specifically, 4$\mu m$ wavelength) where there is
an atmospheric transparency window. Similar consideration can be applied to any other wavelength.

The paper is organized as follows. In section \ref{structure} we present the specifics of photonic structure with phase-changing defect based on vanadium
dioxide. In section \ref{freq} we briefly present a semi-qualitative study of the transport characteristics of an incident monochromatic field in the steady-
state regime. Our main focus is on numerical study of the case of short input pulses; it is presented in section \ref{time}. The modeling of the full transient
electromagnetic pulse propagation coupled with heat transfer analysis is presented in subsection \ref{Atime}, while the analysis of our simulations is 
presented in subsection \ref{Btime}. A generalization for the case of multiple defects is considered in section \ref{Multi}. The results are summarized 
in the final section \ref{conc}.


\section{Photonic crystal with phase changing defects}

\label{structure}

\begin{figure}[h]
\centering\includegraphics[width=9cm]{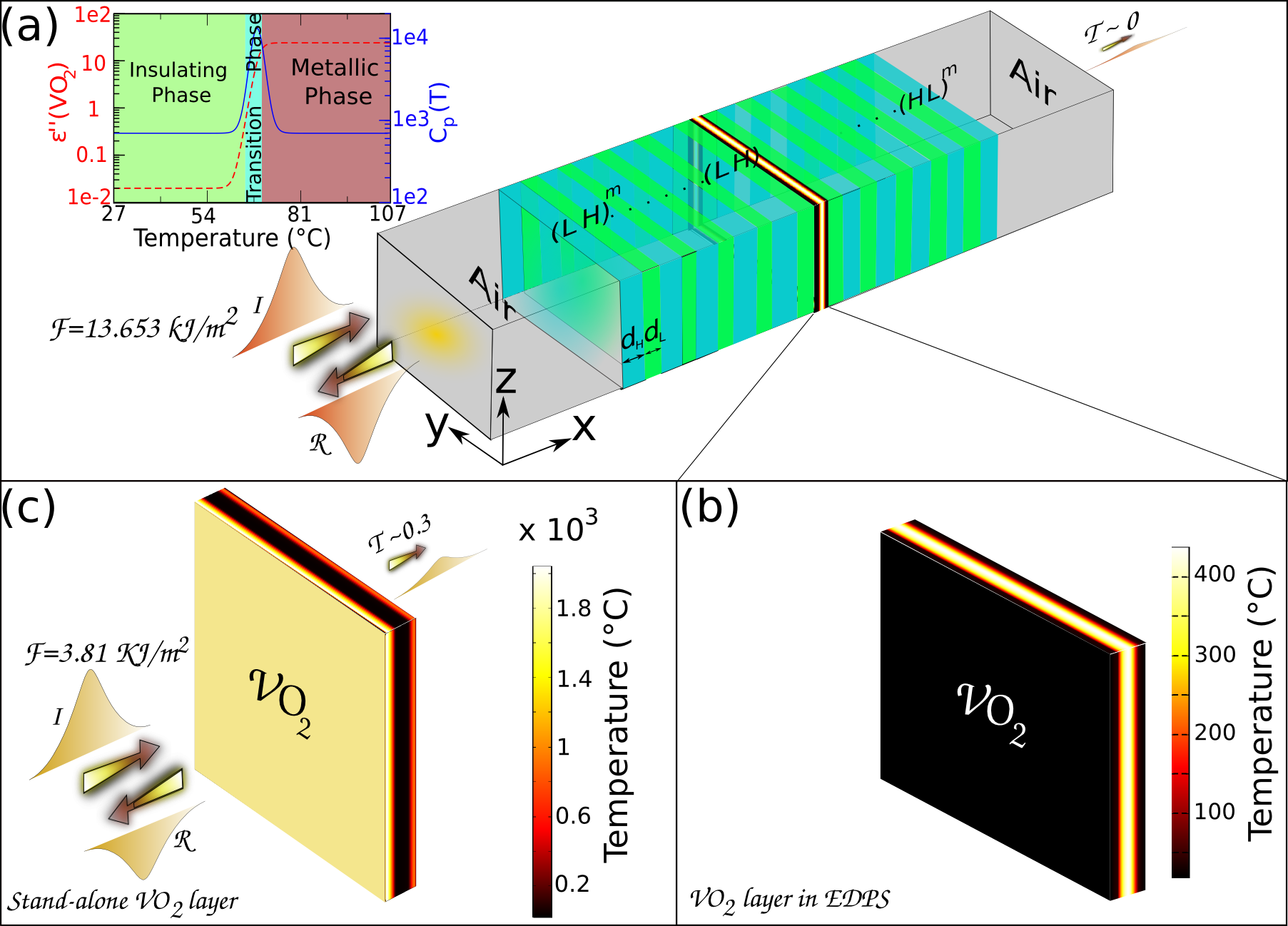}
\caption{(a) Schematic illustration of our proposed EDPS with one defect. It consists of two alternating composite layers with low $L$ ($SiO_2$) and high $H$ ($Si_3N_4$) refractive indices and a defect ($VO_2$) layer
placed at the plane of mirror symmetry of the structure. The two composite
layers have a quarter wavelength width while the defect layer has a width of
half-wavelength. Inset: For low temperatures $VO_2$ remains in the
dielectric phase while for high temperatures it undergoes a phase transition
to a metallic phase. During the phase transition the imaginary part of the
permitivity $\protect\epsilon_D^{\prime\prime}$ changes by three to five
orders of magnitude. A typical behavior of $\protect\epsilon_D^{\prime
\prime}(T)$ is shown in the inset (red line) using Eq. (\protect\ref{fit})
with $\protect\epsilon^{\prime\prime}_0\approx 0.02$, $\Delta\protect\epsilon_D^{\prime\prime}\approx 25$ and 
$\Delta\approx 1^o C$~\protect\cite{Yoon,Chu}. The blue line shows the temperature variation of the specific
heat $C_P$ of the $VO_2$ layer due to the released latent heat during the
dielectric to metal phase transition, see Eq. (\protect\ref{latent}). (b) Magnified view of the thermal density distribution within the $VO_2$ layer in the EDPS for a high intensity input light with $\mathcal{F}$ = 13.653$\frac{kJ}{m^2}$. (c) The thermal density distribution within the $VO_2$ layer in SA structure for an input light with $\mathcal{F}$ = 3.81$\frac{kJ}{m^2}$. In this case the SA already surpasses the melting point of the $VO_2$ layer.}
\label{fig1}
\end{figure}

Consider a multilayer consisting of two alternating quarter-wavelength
dielectric layers $L$ and $H$ having low and high refractive indices
respectively. A PCM defect layer $D$ of half-wavelength thickness $d$ is
embedded at the mirror symmetry plane of the multilayer structure. In the more
general case of multiple defects, their positions will be appropriately
chosen in a way that the total multilayer photonic structure will still respect a
mirror symmetry. Below we shall refer to the set-up which involves embedded defects 
in the photonic structure as EDPS. 
An example of such photonic structure is a layer structure described by
the sequence $(LH)^{m}D\left[ (HL)^{n}LD\right] ^{k-1}(HL)^{m}$, where $k=1,2,\cdots $, correspond to one, two, three etc. defect layers. The
integers $m$ and $n$ denote the number of alternating bilayers $LH$ located
at the two ends and between the defect layer(s). A schematic illustration of
our proposed design for the case of a single-defect (i.e., $k$ = 1) embedded
in a layer structure consisting of $m=5$ bilayers $LH$ is shown in Figure~\ref{fig1}.

In our simulations below, we considered that the layer $L$ consists of
silicon dioxide (SiO$_2$) while the layer $H$ consists of silicon nitride (Si$_3$N$_4$) with respective (real) permittivity 
values of $\epsilon^{\prime}_L=1.9396$ and $\epsilon^{ \prime}_H=5.7312$. We have
further assumed that at the operational wavelength of 4 $\mu$m (mid-wave
IR), these materials have negligible absorption i.e. $\epsilon_{L/H}^{\prime\prime}\approx 0$. The defect layer consists of a PCM that undergoes a phase
transition from a dielectric to a metallic phase as the temperature increases, due to
heating from the incident light. As an example we consider
vanadium dioxide ($VO_2$) which has been studied most intensively because of
its near-room- temperature phase transition as well as of its high phase
stability. It undergoes a first-order metal-to-dielectric transition (MDT) 
around $T_C\approx 68^oC$ from a high temperature metallic phase to a low
temperature dielectric phase which is accompanied by a structural phase
transition from a high temperature rutile to a low temperature monoclinic
structure. As a result of the MDT, a dramatic change in its complex
permittivity $\epsilon_D=\epsilon^{\prime}_D+i\epsilon^{\prime\prime}_D$
occurs.

Although various experimental studies for $VO_2$ give different values of
its perimittivity in the dielectric and the metallic phase, all of them seem
to agree that the changes in $\epsilon_D^{\prime\prime}$ can be as high as
four (even five) orders of magnitude in the MIR domain~\cite{Morin, Averitt12, Kats13,
Berglund68, Basov07}. The variation in the exact values is mainly attributed
to the preparation and the quality of the $VO_2$ sample and in the
experimental techniques used to measure its permittivity. In the simulations
below we will consider this abrupt change in $\epsilon^{\prime\prime}_D$
near the phase transition temperature $T_C$. We shall therefore model the
temperature $T$-dependent imaginary permitivity of $VO_2$ with the function 
\begin{equation}  \label{fit}
\epsilon^{\prime\prime}_D(T)=\epsilon^{\prime\prime}_0 +\left[{\frac{\Delta\epsilon^{\prime\prime}_D}
{\exp\left[-(T-T_C)/\Delta\right]+1}}\right].
\end{equation}
The constants $\epsilon^{\prime\prime}_0$ and $\Delta\epsilon^{\prime\prime}_D$ have been extracted from various experimental data presented in
the literature in a way that matches the observed values of $\epsilon_D^{\prime\prime}$ in the dielectric and metallic 
phase ~\cite{Kats13, Basov07, Atwater, Dressel16,Kana}. We have 
found that $\epsilon^{\prime \prime}_0$ can take values between $0.02$ (or even lower)
to $0.1$ while the extracted $\Delta\epsilon^{\prime\prime}_D$ varies
between $25-80$. We have performed simulations using two set of parameters $(\epsilon_0^{\prime\prime},\Delta \epsilon^{\prime\prime}_D)$ corresponding
to $(0.1, 25)$ (configuration I) and $(0.02, 25)$ (configuration II). A
dependence of $\epsilon^{\prime\prime}_D(T)$ on temperature $T$ is shown in
the inset of Fig. \ref{fig1} (dash lines) for configuration II. In both cases the
results for the transport characteristics of our structure remain qualitatively the same. The "smoothing" parameter $\Delta$
over which the transition occurs is assumed to be $\Delta\approx 1^o C$ (see
section \ref{time}). Nevertheless, we have checked that other values of 
$\Delta$ up to $4^o C$ show the same qualitative picture of the transmission
properties of our photonic structure. Finally for simplicity we have assumed
that the real part of permittivity remains approximately constant $\epsilon^{\prime}_D\approx 8.41$~\cite{Gavinit}.


\section{Frequency domain analysis}

\label{freq}

Let us start with the semi-qualitative frequency domain analysis of
steady-state regime. It will provide us with the understanding of what to expect
from the time-domain simulations for the case of short pulses presented in
the next section.

The electric component of a time-harmonic monochromatic field satisfies the
Helmholtz equation 
\begin{equation}
\nabla ^{2}{\vec{E}}(x)+{\frac{\omega ^{2}}{c^{2}}}\epsilon (x){\vec{E}}(x)=0
\label{Helm}
\end{equation}
where $\epsilon (x)$ is the spatially varying permittivity of the structure
along the propagation direction $x$, $c$ is the speed of light in the vacuum
and $\omega $ is its frequency. The scattering fields, and therefore the
frequency-dependent transmittance $\mathcal{T}$, reflectance $\mathcal{R}$,
and absorption $\mathcal{A}$ have been calculated numerically from Eq. (\ref{Helm})

The frequency dependent transmittance $\mathcal{T}(\omega )$, reflectance $\mathcal{R}(\omega )$ 
and absorption $\mathcal{A}(\omega )$ are defined as: 
\begin{equation}
\mathcal{T}(\omega )={\frac{S_{\mathrm{tr}}(\omega )}{S_{\mathrm{in}}(\omega
)}},\,\mathcal{R}(\omega )={\frac{S_{\mathrm{refl}}(\omega )}{S_{\mathrm{in}}(\omega )}},
\,\mathcal{A}(\omega )=1-\mathcal{T}(\omega )-\mathcal{R}(\omega ),  \label{CW}
\end{equation}
where $S(\omega )={\frac{1}{2}}\mathcal{R}e\left( {EH}^{\ast }\right)$
is the real-valued energy flux normal to the layers. Using Eqs. (\ref{Helm}, \ref{CW}) we can calculate the 
transmission spectrum for the case of one $k=1 $ defect embedded in a photonic crystal with $m=5$ bilayers.

\begin{figure}[h]
\centering\includegraphics[width=9cm]{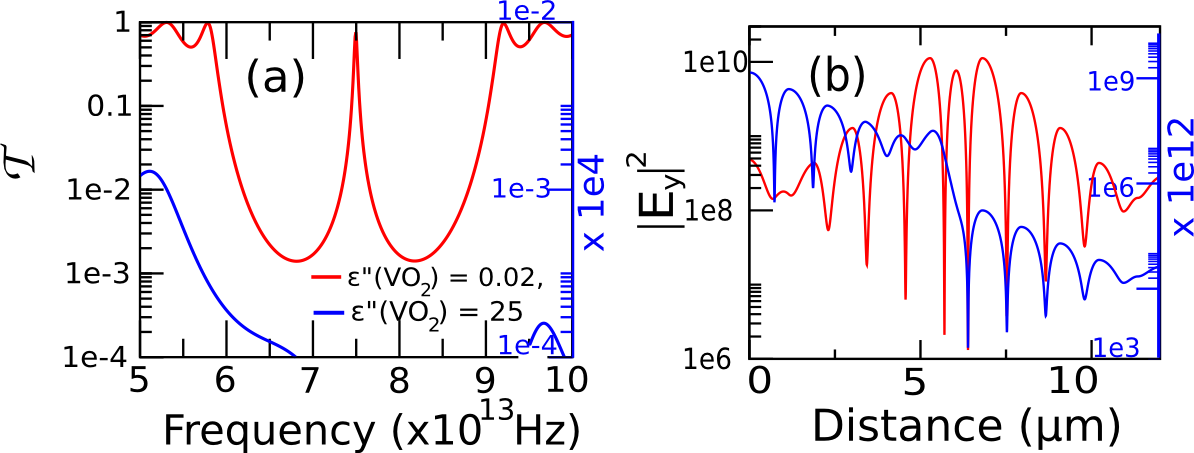}
\caption{(a) Transmission spectra obtained for the single VO$_2$ layer in our $EDPS$ of Fig. \protect\ref{fig1}. Red lines correspond to $\protect\epsilon^{\prime
\prime}_D = 0.02$ (dielectric phase) and blue lines correspond to $\protect\epsilon^{\prime\prime}_D = 25$ (metallic phase) [configuration 
II associated with Eq. (\protect\ref{fit})]. (b) The associated resonant defect mode profiles (colors as in (a)). When the fluency of the incident 
signal is low, the $VO_2$ layer is not heated up and remains in the dielectric phase with $\protect\epsilon^{\prime\prime}_D \approx \protect
\epsilon^{\prime\prime}_0=0.02$. In this case the structure supports a defect resonant mode which is exponentially localized within the defect 
layer. When the incident signal has large fluency, the $VO_2$ layer is heated up and its imaginary permitivity increases dramatically i.e. $\protect
\epsilon^{\prime\prime}_D$= 25 (see inset of Fig. \protect\ref{fig1}). In this case the defect resonant mode is destroyed and the electric field 
intensity at the position of the defect layer decreases exponentially from its incident value.}
\label{fig2}
\end{figure}

At low light intensity, the heating of the VO$_{2}$ layer is negligible, and
thus its temperature $T$ remains below the critical phase transition temperature value $T_{C}$. In this
case, the VO$_{2}$ defect layer is in the dielectric state with the low value
of $\epsilon _{0}^{\prime \prime }$. The low-intensity transmittance as a
function of frequency is shown in Fig. \ref{fig2}a in red. The resonant
transmission peak in the figure corresponds to the frequency of the
localized defect mode. The resonant mode profile is shown in Fig. \ref{fig2}b, also in red.
If the light intensity is high enough, the VO$_{2}$ layer is heated above
the transition temperature $T_{C}$. In this case, the VO$_{2}$ defect layer is in
the metallic state indicated by the large value of $\epsilon _{D}^{\prime \prime }$.The latter results in suppression of the localized mode, along with the
resonant transmission, as seen in Fig. \ref{fig2}a and b (see the blue lines).

A simple way to explain the destruction of the resonant transmission via the
defect mode is by realizing that its existence is a result of two competing
dissipation mechanisms. The first one is the effect of losses due to
radiation from the boundaries of the layered structure. An estimation of their magnitude is
given from the linewidth $\Gamma _{\mathrm{R}}$ of the associated resonant
mode. The second one is the Ohmic losses $\Gamma _{\mathrm{Ohm}}$ occurring
in the high-temperature (metallic) phase of $VO_{2}$. One can estimate these
losses to be: 
\begin{equation}
\Gamma _{\mathrm{R}}\sim {\frac{1}{\xi }}\exp (-2L/\xi );\quad \quad \Gamma
_{\mathrm{Ohm}}\sim {\frac{\omega }{c}}\epsilon _{D}^{\prime \prime }{\frac{d}{\xi }}  \label{Rloss}
\end{equation}
where $L$ in Eq. (\ref{Rloss}) is the total thickness of the layered structure. In the {\it under-damping} limit $\Gamma _{\mathrm{R}}\gg 
\Gamma_{\mathrm{Ohm}}$, photons can resonate via the high-Q resonant defect mode through the layer structure, thus leading to high transmittivity
i.e. ${\cal T}(\omega_r)\approx 1$. In this limit both reflection and absorption are vanishing i.e. ${\cal R}(\omega_r)\approx 0, {\cal A}
(\omega_r)\approx 0$. As $\epsilon _{D}^{\prime\prime }$ increases, the absorption coefficient $\mathcal{A}$ increases initially linearly  
and reaches a maximum value at some $\epsilon_D^{\prime\prime CC}$ with a simultaneous decrease (increase) in $\mathcal{T}$ ($\mathcal{R}$) ($CC$ denotes the critical coupling). 
The maximum absorption occurs when {\it critical coupling} is achieved i.e. when Ohmic and radiative losses are optimally balanced. In the 
domain $\epsilon_D^{\prime\prime}<\epsilon_D^{\prime\prime CC}$ the dominant dissipation mechanism is still associated with radiative 
losses and the incident photons spend enough time in the resonant mode in order to be eventually absorbed, thus increasing the ohmic 
losses at the $VO_{2}$ layer. A further increase in $\epsilon _{D}^{\prime \prime} $ leads to a noticeable decrease in $\mathcal{A}$ and 
a simultaneous increase (decrease) in reflection (transmittance). This is the {\it over-damping} limit $\Gamma_{\mathrm{R}}\ll \Gamma_{
\mathrm{Ohm}}$ where the dominant dissipation mechanism is associated with the Ohmic losses. The associated dwell time is 
$\tau_{\mathrm{dwell}}^{-1}\sim \Gamma_{\mathrm{Ohm}}$ -- thus the incident photons do not dwell for long enough time inside the 
cavity in order to be absorbed by the $VO_{2}$ layer. Rather, they are reflected back to space, thus increasing the reflectivity of the photonic 
structure. For even larger values of $\epsilon_{D}^{\prime \prime }$, a complete destruction of the resonant mode takes place 
and the entire energy is reflected i.e. $\mathcal{R}\approx 1$ while the absorbed energy is essentially zero $\mathcal{A}\approx 0$. In 
simple terms, the $VO_{2}$ defect, being in the metallic phase, has completely decoupled the left and right parts of the 
photonic crystal which now acts as a (almost) perfect mirror.


\section{Short pulses}

\label{time}

We now turn to the study of the transport characteristics of the layered structure of Fig.~\ref{fig1} in the case of short-pulse incident beams. In
subsection \ref{Atime} we will present the mathematical model that describes the propagation of a beam in the presence of a temperature
-varied defect permittivity. The results of the time-domain simulations and the conclusions about the transport characteristics of the layered structure, 
in the case of short incident pulses, will be presented in the subsequent section \ref{Btime}.


\subsection{Time-domain electromagnetic and heat-transfer model set-up}

\label{Atime}

The pulse propagation in the case of temperature-varying permittivity of the $VO_2$ layer is described by the following set of coupled electromagnetic
and thermal equations: 
\begin{subequations}
\label{subeqns}
\begin{align}
\bigtriangledown\times\vec{H} = \vec{J} + \epsilon^{\prime}(x)\frac{\partial\vec{E}}
{\partial t},\quad \bigtriangledown\times\vec{E} = -\mu(x)\frac{\partial{\vec{H}}}{\partial t},  \label{subeq1} \\
\rho_D C_{p}^D\frac{\partial T}{\partial t} -\nabla.(k_D\nabla T) =Q,\quad
Q= \vec{J}\!\cdot\vec{E},  \label{subeq2}
\end{align}
where $\mu(x)=\mu_0$ is the permeability of the composite materials, ${\vec J}=\sigma(T) {\vec E}$ is the current density, $k_D$ is the thermal
conductivity, and $\sigma(T)$ is the electrical conductivity which in the case of the $VO_2$ layer(s) changes as a function of temperature $T$
(measured in degree Celsius). The latter can be extracted using Drude's model from the expression for the imaginary permittivity $\epsilon_D^{
\prime\prime}(T)$ at $4\mu m$ wavelength (we do not consider dispersion effects here), see inset of Fig. \ref{fig1}. In the above coupled 
electromagnetic-heat transfer model, the absorbed power $Q$ (in units of W\!/\!m$^3$) deposited per unit volume within the defect layer(s) from the incident electromagnetic pulse
lead to an increase of temperature $T$ within the $VO_2$ layer as described
by the rate equation (\ref{subeq2}). The parameters $\rho_D, C_p^D$, and $k_D $ correspond to the mass density (kg\!/\!m$^3$), specific heat capacity at
constant pressure (J/kg K), and thermal conductivity (W\!/\!m K) of the $VO_2$
layer. At the first order MDT the value of $C_p^D$ will change sharply due
to the released latent heat which has been measured to be $H_L\approx 5.042
10^4 J/kg$ \cite{LeClair, Berglund69}. The accommodated latent heat $\Delta
Q_L$ is then assumed proportional to the conductivity change i.e. $\Delta Q_L={\frac{H_L}{\Delta \sigma_t}}{\frac{d\sigma}{dT}} dT$. As a result, the specific
heat capacity of $VO_2$ is 
\end{subequations}
\begin{equation}  \label{latent}
C_p^D(T)=C_p^{(0)}+{\frac{H_L}{\Delta \sigma_t}} \frac{d\sigma}{dT}
\end{equation}
where $\Delta \sigma_t\sim 9.96 \cdot 10^4$ (S/m) is the total conductivity jump
during the phase transition and $C_p^{(0)}\approx 700 J kg^{-1} K^{-1}$ is
the specific heat capacity (which is assumed to remain approximately
constant). The model dependence of $C_p^D(T)$ Eq.~(\ref{latent}) on
temperature for the parameters $\epsilon_0^{\prime\prime}=0.02$, $\Delta
\epsilon^{\prime \prime}=25, \Delta=1^oC$ associated with configuration II
(see discussion below Eq.~(\ref{fit})) is reported in the inset of Fig. \ref{fig1} with a blue solid line. A similar temperature dependence of $C_p^D(T)$ is
found for configuration I as well (not shown). Finally in our calculations
we have also considered the changes in thermal conductivity $k_D $ occurring during the phase transition. Specifically we have assumed
that $k_D=4 \frac{W}{mK}$ in the dielectric phase while it takes the value $k_D=6 \frac{W}{mK}$ in the metallic phase~~\cite{Ramakrishna}. The Maxwell's
equations (\ref{subeq1}) have been solved together with the conductive heat
transfer equation (\ref{subeq2}), using a commercially available Comsol
Multiphysics software~\cite{comsol}.

In all our simulations we have considered Gaussian modulated incident pulses with a carrier frequency of $f_0 \approx$ 7.5e13 Hz (wavelength $\lambda_0$ 
= 4 $\mu$m). Note that this frequency corresponds to the defect resonant mode supported by the layered structure of Fig. \ref{fig1} (shown in the transmission 
spectra in Fig. \ref{fig2}a). The associated time-varying electric field $E(t)$ of the incident pulse is: 
\begin{equation}
E(t) = E_0\exp{\left[-0.5(t/\tau)^2\right]}\cos(k_0x-2\pi f_0t)  \label{eq:2}
\end{equation}
where $E_0$, $t$, $\tau$ and $k_0$ denotes the peak electric field
amplitude (V/m), time (s), width (s) and wave vector (2$\pi$/$\lambda_0$) of the input signal, 
respectively. The minimum pulse width $\tau$ was chosen in such a way that for low incident fluences the initial
packet has a maximum spectra overlap with the transmission spectrum of the
photonic structure near the resonant defect mode, see Fig. \ref{fig2}. This
requirement guaranties a maximum transmission for low fluencies. The accuracy of the time integration was checked
by decreasing the time-step by half and confirmed that the results of the simulations remain unchanged.

For a given $\tau$ and $E_0$, the energy fluence was evaluated as $\mathcal{W}=(\sqrt\pi/2$)c$\epsilon_0 E_0^2 \tau$. In the simulations the pulse width 
$\tau$ was smaller than $0.5 nsec$ due to numerical integration constrains. Nevertheless one can extrapolate the validity of our results for pulse durations as 
large as the thermal relaxation time $\tau_D\approx l^2/D\sim 10 nsec$ where $D=k/\rho c_p$ is the thermal diffusion constant and $l\sim0.13 \mu m$ is 
the thermal relaxation length. The latter was estimated from our simulations by evaluating the FWHM of the temperature profile inside the $VO_2$ layer when 
it reaches its maximum value (i.e. for $\mathcal{F}$ = 13.653 kJ/m$^2$).

Using the above calculation scheme we have evaluated the total $\mathcal{T}$, $\mathcal{R}$ and $\mathcal{A}$ numerically using the expressions: 
\begin{equation}  \label{ttran}
\mathcal{T}= \frac{\bigintss_0^{\infty} {\bar P}_{\mathrm{tr}}(t) dt} {\bigintss_0^{\infty} {\bar P}_{\mathrm{in}} (t)dt},\quad \mathcal{R}= 
\frac{\bigintss_0^{\infty} {\bar P}_{\mathrm{refl}} (t)dt}{\bigintss_0^{\infty} {\bar P}_{\mathrm{in}} (t)dt}, \quad \mathcal{A}= 1 - \mathcal{T}- \mathcal{R},
\end{equation}
where ${\bar P}_j(t)$ is the energy power flowing across the incident (j=in, refl) and exit surfaces (tr) of the
structure.


\subsection{Simulation result and discussion}

\label{Btime}

\begin{figure}[h!]
\centering\includegraphics[width=8cm]{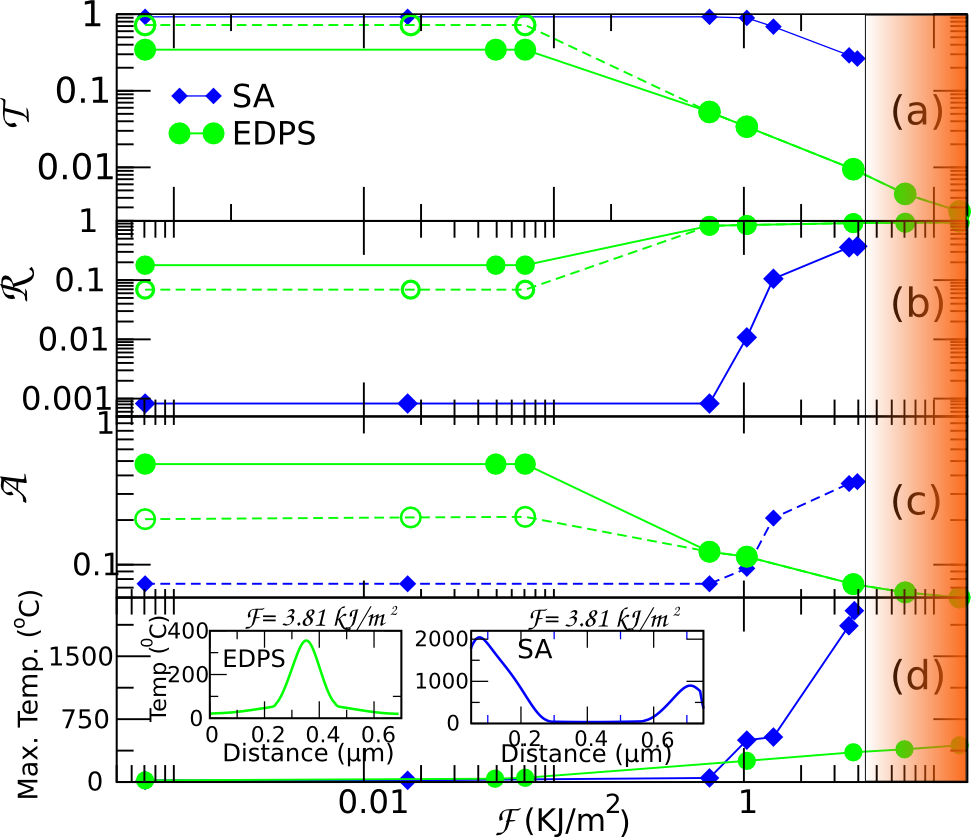}
\caption{The time-cumulative (a) transmittance ($\mathcal{T}$); (b) reflectance ($\mathcal{R}$); and (c) absorption ($\mathcal{A}$) versus the fluency 
$\mathcal{F}$ of an incident pulse. We consider the case of the VO$_2$ layer in EDPS, of Fig. \protect\ref{fig1}. The EDPS demonstrates high transmittance 
for pulses with lower fluency values while it becomes completely reflective for pulses with high fluency values. The filled circles correspond to parameters 
$\protect\epsilon_0^{\prime\prime}\approx 0.1, \Delta\protect\epsilon_D^{\prime \prime}=25, \Delta=1^oC$ (configuration I) while the empty circles 
to $\protect\epsilon_0^{\prime\prime}\approx 0.02, \Delta \protect\epsilon_D^{\prime\prime}=25,\Delta=1^oC$ (configuration II). The transmission 
characteristics of the layered structure is compared with that of a stand alone $VO_2$ layer (blue filled diamonds) where a significant absorption occurs 
for high fluency pulses. In (d) we report the maximum temperature reached inside the $VO_2$ layer for incident pulses with various fluencies $\mathcal{F}$. 
In the insets we show the spatial distribution of the temperature inside the $VO_2$ layer. The highlighted orange area indicates the fluency regime for 
which the SA $VO_2$ layer has reached temperatures higher than its melting temperature $T\approx 1967^oC$.}
\label{fig3}
\end{figure}

A summary of our time-domain simulations for $\mathcal{T}, \mathcal{R}, \mathcal{A}$ Eq. (\ref{ttran}) for the case of a single $VO_2$ defect layer, 
embedded in a multilayer photonic structure (see Fig.~\ref{fig1}), versus the fluency $\mathcal{F}$ of the incident pulses is shown in Fig.~\ref{fig3} 
(open/filled green circles). Filled (configuration I) and open (configuration II) symbols correspond to two different values of $\epsilon_0^{\prime\prime}$
~\cite{Atwater, Kana}, see section \ref{structure}. At the main part of Fig. \ref{fig3}d we present a summary of the maximum temperatures reached 
inside the $VO_2$-layer during the simulations for pulses of different fluencies $\mathcal{F}$. In our simulations we have also monitored the temperature 
profile within the $VO_2$ layer. Some typical spatial temperature profiles are reported in the insets of Fig. \ref{fig3}d for times at the end of the simulation 
pulse. We find that in the case of the stand-alone $VO_2$ layer the temperature profile reached its maximum value near the boundary interfaces. In 
contrast, for the EDPS the maximum temperature is reached at the center of the defect where the electric field experiences its maximum value (see 
Fig.~\ref{fig2}b). At the same figure we report for comparison purposes the transport characteristics of a SA $VO_2$ layer (blue filled diamonds).

Let us discuss in more detail the transport features of the layered structure. Incident pulses with fluency $\mathcal{F}< 0.07 kJ/m^2$ do not cause any 
significant heating (see Fig. \ref{fig3}d) of the defect layer, resulting in a negligible increase in $\epsilon^{\prime \prime}_{D}(T)$. In this case the
resonant defect mode remains unaffected (see discussion in section \ref{freq}) and dictates the transport characteristics of the incident pulse.
Specifically we find that the transmittance can be larger than 35-40\% for $VO_2$ films with $\epsilon_0^{\prime\prime}\approx 0.1$ (configuration 
I) in the dielectric phase, see filled symbols in Fig. \ref{fig3}. For $\epsilon_0^{\prime\prime}\approx 0.02$ (configuration II) the transmittance
can reach values as high as 75\%, see open symbols in Fig. \ref{fig3}.

As the fluency of the pulse increases beyond $\mathcal{F}\geq 0.07kJ/m^2$, the induced heating effects become important. Specifically the 
temperature at the $VO_2$ layer goes above $T_C$ driving the defect layer to the metallic phase. In all our simulations we have found that the 
time duration for which the permittivity $\epsilon_D^{\prime\prime}(T>T_C)$ changes significantly is typically less than 0.01 ps. At this point 
the resonant defect mode has been completely suppressed (see Figs. \ref{fig2}a,b) leading to a sharp decrease in transmittance, see Fig. \ref{fig3}a. 
In contrast, the transmission of the $VO_2$ SA layer is still $\mathcal{T}\approx 1$ for fluencies as high as $\mathcal{F}\approx 1kJ/m^2$.  
The simulations of Fig. \ref{fig3}a indicate that the input fluency threshold (IFT)  of our layer structure can be at least one order of magnitude smaller 
than the input fluency threshold provided by the stand-alone $VO_2$ layer. The low value of $\mathcal{F}_{\mathrm{IFT}}$ of the layer structure is another
consequence of the exponential sensitivity of the transition to the over-damping regime due to the nature of the resonant defect mode (see 
Eq. (\ref{Rloss}) and relevant discussion). The under-damping-to-over-damping transition is also responsible for the increase in reflectance which 
for fluencies $\mathcal{F}\geq 1kJ/m^2$ is approximately 100\%. At the same time the absorptivity $\mathcal{A}$ decays to zero (see discussion in
section \ref{freq}). This behavior has to be contrasted with the stand alone $VO_2$ layer, which has already reached temperatures above the
melting point $T_{\mathrm{melt}}\approx 1967^oC$ for fluencies $\mathcal{F}\approx 4 kJ/m^2$ (see Fig. {\ref{fig3}d)~\cite{Cheng}. We therefore
conclude that the EDPS design provides to the $VO_2$ layer a protection from overheating and thus prevents its destruction from high fluency 
incident radiation. 

The EDPS of Fig. \ref{fig1}a can be utilized as a photonic limiter. These are devices that are used to protect sensitive sensors by blocking high fluency 
(or power) incident radiation while transmitting low level radiation. Many of the existing limiters achieve this goal by absorbing most of the energy of
the incident pulse -- an operation that leads to their overheating and destruction (sacrificial limiters). A typical example of such limiter is the SA 
$VO_2$ layer. Instead, the layered structure of Fig. \ref{fig1}a has an improved \textit{damaged threshold} (DT) since it reflects back to space the harmful high fluency 
radiation. At the same time it demonstrates a low $\mathcal{F}_{\mathrm{IFT}}$ which is considered another important feature of an efficient photonic limiter. 
It is usually refered as the {\it limiting threshold} of the limiter and it has always to be well below the damage threshold of the sensor that the limiter
is protecting. The ratio between these two thresholds provides the figure of merit of the limiter and it is known as its \textit{dynamical range} (DR). Our 
numerical simulations clearly demonstrate that the EDPS limiter has increased its DR by at least two orders of magnitude as compared to the
SA layer. 


\section{Generalization to Multiple Defects}

\label{Multi}

Next we generalize the above study for the case of multiple $VO_2$ defects. The transmission spectra for three ($k=3$) and eight ($k=8$) defects 
are reported in Figs. \ref{fig4}a,b while the profiles of the associated defect modes are shown in Figs. \ref{fig4}c,d. These calculations have been
done for low light intensities (red lines) and high light intensities (configuration I, see section \ref{structure}). The new element in these multi-defect 
configurations (as opposed to the one defect case, see Fig. \ref{fig2}) is the appearance of other resonant modes in the vicinity of the center of the 
band-gap which lead to the creation of a mini-band (see red lines in Figs. \ref{fig4}a,b). 

\begin{figure}[h]
\centering\includegraphics[width=8cm]{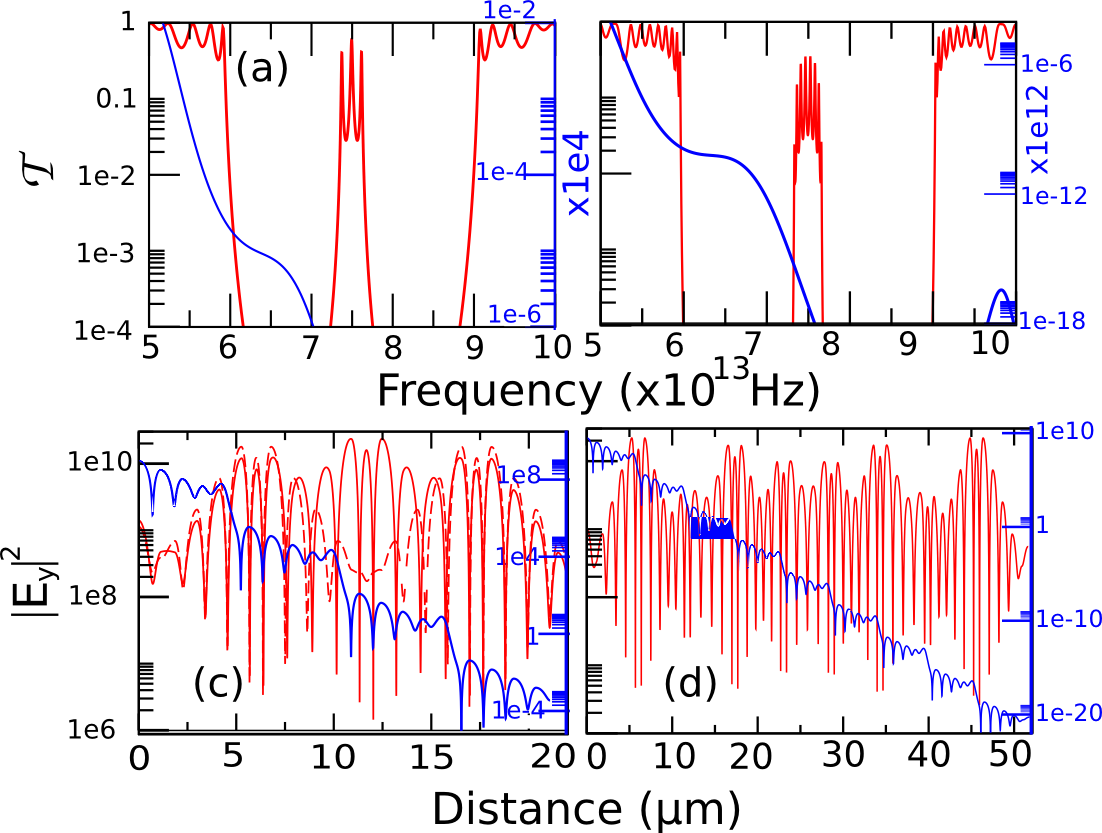}
\caption{Transmission spectra obtained for the layered structure incorporating (a)
three $VO_2$ defects, and (b) eight $VO_2$ defects. The defects are placed
in a way that the whole structure respects a mirror-symmetry. Two scenarios
corresponding to $\protect\epsilon^{ \prime\prime}_D$ = 0.02 (dielectric
phase) and 25 (metallic phase) of the $VO_2$ layer are depicted with red and
blue solid lines respectively. (c) and (d) show the associated resonant defect mode profiles for the case of the EDPS incorparating three $VO_2$ defect and (d) eight $VO_2$ defects respectively. The red dash and solid lines in (c) denote the symmetric and asymmetric mode profiles at the resonant frequencies of 7.373e13 Hz and 7.495e13 Hz for the case when $\protect\epsilon^{ \prime\prime}_D$ = 0.02.}
\label{fig4}
\end{figure}

We understand the formation of the mini-band as follows. Consider identical single-defect layer structures. From the discussion at section \ref{freq} we know that
each of these configurations can support a defect mode which is exponentially localized around the defect. When these identical singe-defect layer structures are 
brought in some finite distance $L_d$ from one another (i.e. when the portion of the layer structure which is surrounding the defects becomes finite), the defect 
mode associated with each configuration can be evanescently coupled with its degenerate pair supported by the nearby configuration. The coupling
constant between them is $q\sim \exp(-L_d/\xi)$. This coupling lifts the degeneracy and leads to the creation of the mini- bands (within the band-gap
of the combined layer structure) with bandwidth proportional to $q$. The associated states are symmetric and antisymmetric linear combinations of the single
defect modes and their profiles have a multi-humped shape with each hump located in the neighborhood of a defect (see Figs. \ref{fig4}c,d for a three
and an eight defect mode configuration). The above picture is valid as long as the total losses (i.e. the sum of the radiative losses due to leakage and
the ohmic losses due to small $\epsilon_D^{\prime\prime}$ at the defects) are much smaller than the coupling $q$. This requirement is satisfied as
long as the incident field carries a small fluency -- so that the $VO_2$ defects are in the dielectric phase where $\epsilon_D^{\prime\prime}\approx
\epsilon_0^{\prime\prime}\approx 0$. Similarly to the one defect case, in this limit the transport is dictated by the presence of multi-hump mirror
symmetric states which lead to high transmittivity $\mathcal{T} \approx 1$ (see red solid lines in Figs. \ref{fig4}a,b).

Once the incident fluency increases, the temperature at the defect layers increases above the MDT value $T_C$. The Ohmic dissipation at the $VO_2$ 
layers increases dramatically and the total losses are larger than the coupling $q$. In this case the same scenario as in the case of one defect applies; 
namely the multi-humped modes are destroyed (see blue lines in Figs. \ref{fig4}c,d), leading to a suppression of transmittance $\mathcal{T}\approx 0$ 
(blue lines in Fig. \ref{fig4}a,b) and a consequent increase/decrease of reflectance/absorption i.e. $\mathcal{R}\approx 1$ and $\mathcal{A}\approx 0$ 
respectively.

In Figs. \ref{fig5}a,b,c we report the transport characteristics of our layered structure for the case of short incident pulses for $k=3$. In this case the advantage 
of the mini-band formation is translated to a tolerance in choosing a central pulse frequency anywhere inside the mini-band i.e. $f_0\pm q$ and, 
nevertheless, have high transmittivity at low fluencies. At the same time these pulses will be (almost) completely reflected when their fluency is high enough. 
The underlying mechanism is the same as the one found for the case of a layered structure with a single $VO_2$ defect (see previous section) and it is associated with 
the light-induced phase transition of the  $VO_2$ layer and its modifications in the presence of a layer structure.

\begin{figure}[h]
\centering\includegraphics[width=9cm]{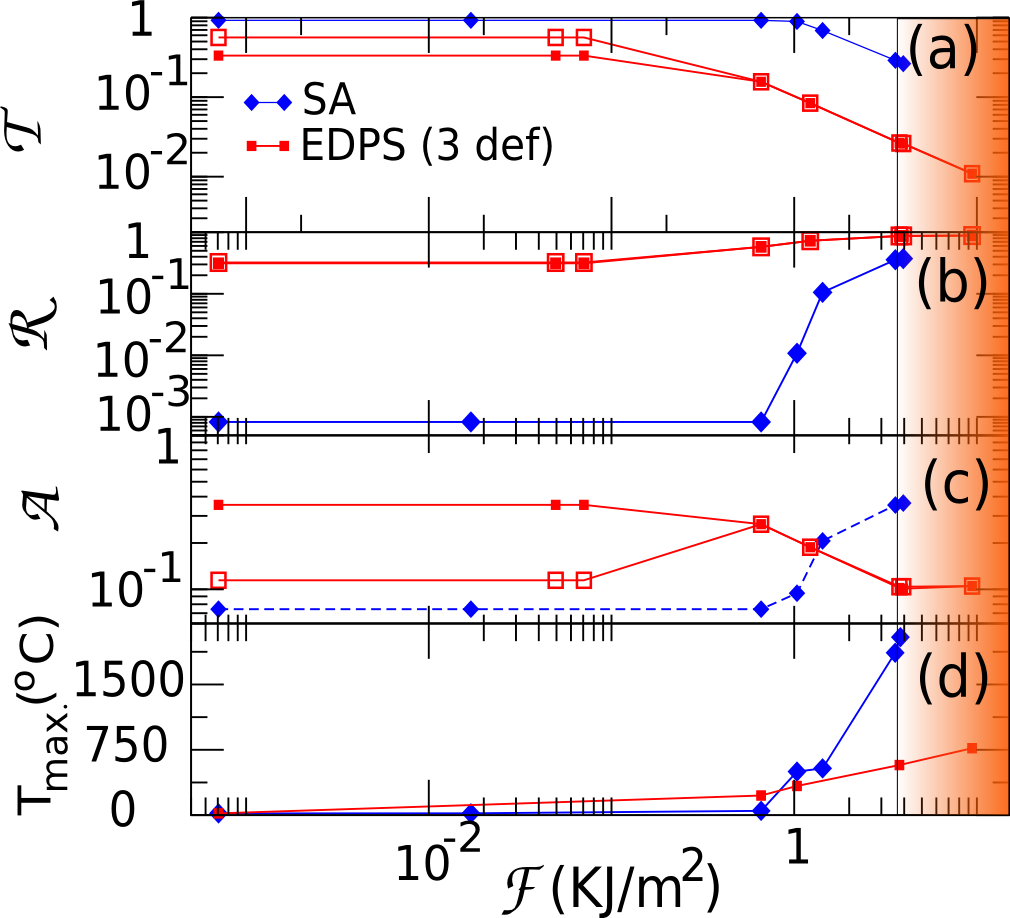}
\caption{Time-cumulative (a) transmittance ${\cal T}$; (b) reflectance ${\cal R}$; and (c) absorption ${\cal A}$ versus the fluency $\mathcal{F}$ for the case of three ($k=3$) $VO_2$ defects embeded in a multilayered structure. (d) The maximum temperature reached inside the 
$VO_2$ layer during the simulation period versus the fluency of the incident pulse. The orange highlighted domain correspond to fluencies for which the SA $VO_2$ layer has already reached its melting temperature of $1967^oC$. In all the sub-figures filled symbols indicate simulations with permittivity parameters associated with configuration I while open symbols indicate simulations with permittivity parameters associated with configuration II.}
\label{fig5}
\end{figure}


\section{Conclusion}

\label{conc}

In summary, let us highlight the most important features of the transmission properties of the planar resonant cavity in Fig. \ref{fig1} containing a
PCM represented by VO$_{2}$. If the input pulse duration is much greater than the thermal relaxation time of the PCM layer, the layered structure
acts as a reflective irradiance limiter described in~\cite{Kottos14}. Specifically, the transition from the resonance transmission to high reflectivity 
associated with the dielectric-to-metal phase transition occurs at much lower input light intensity, compared to the same PCM layer taken out the 
photonic resonant cavity. After the phase transition is completed, the field intensity at the PCM layer becomes much lower than that of the incident 
light, and the entire structure becomes highly reflective, which prevents it from overheating. If the pulse duration is much shorter than the thermal 
relaxation time of the PCM layer, the structure acts as a reflective fluency limiter described in~\cite{Kottos15}. Yet, there is a big difference between 
the approach \cite{Kottos14},\cite{Kottos15} based on optical materials with nonlinear absorption incorporated into a photonic structure, and the 
current approach based on a PCM. In the former case, the transition from high transmittance to high reflectivity may require orders of magnitude 
change in the input light intensity. By contrast, a judiciously designed photonic structure incorporating a PCM can provide an abrupt transition from 
nearly perfect transmittance to nearly perfect reflectance at a desired irradiance or fluence. In either case, using a chain of coupled defect layers 
instead of a single one alleviates the bandwidth restrictions and ensures enhanced performance even for very short input pulses. 

{\it Acknowledgements -}  The authors would like to thank Dr. M. Kats for many useful discussions on phase-change materials. We acknowledge 
support from ONR via grant N00014-16-1-2803 (R. T. \& T. K.), from AFOSR  via grant LRIR14RY14COR (I. M. V.) and from NSF via Grant No. 
DMR-1306984 (F. M. E).

\end{document}